\documentclass[platex,dvipdfmx,fp,twocolumn]{revtex4-1}

\usepackage{amsmath}
\usepackage{amssymb}
\usepackage{multirow}
\usepackage{bm}
\usepackage{newtxmath}
\usepackage{newtxtext}
\usepackage{graphicx}
\usepackage{physics}
\usepackage{algpseudocode}
\usepackage{mathtools}
\usepackage{braket}

\newcommand{\degree}[1]{${#1}^{\circ}$}

\begin{document}
\title{Theory of Tunneling Effect in 1D AIII-class Topological Insulator (Nanowire) Proximity Coupled with a Superconductor}

\author{Ryoi Ohashi, Yukio Tanaka and Keiji Yada}
\affiliation{
Department of Applied Physics, Nagoya University, Nagoya 464-8603, Japan
}

\begin{abstract}
We study the tunneling effect in an AIII-class insulator proximity
coupled with a spin-singlet $s$-wave superconductor, in which 
three phases are characterized by the integer topological invariant 
$\mathcal{N}$.
By solving the Bogoliubov--de Gennes equation explicitly, we analytically obtain
a normal reflection coefficient $R_{\sigma\sigma'}$ and an Andreev 
reflection coefficient $A_{\sigma\sigma'}$, and derive a charge 
conductance formula,
where $\sigma(\sigma')$ is the spin index of a reflected (injected) wave.
The resulting conductance indicates a wide variety of line shapes:
(i)gap structure without coherence peaks for $\mathcal{N}=0$,
(ii)quantized zero-bias conductance peak (ZBCP) with height 
$2e^{2}/h$ for $\mathcal{N}=1$,
and (iii)ZBCP spitting for $\mathcal{N}=2$.
At zero bias voltage $eV=0$,
$\sum_{\sigma\sigma'} R_{\sigma\sigma'} = \sum_{\sigma\sigma'} 
A_{\sigma\sigma'}$ 
is satisfied and the spin direction of an injected electron 
is rotated at approximately \degree{90} 
for the $\mathcal{N}=1$ state. 
Meanwhile, $A_{\sigma\sigma'}=0$ is satisfied for 
the $\mathcal{N}=2$ state, and the spin rotation angle can become \degree{180}.
\\
\end{abstract}

\maketitle

\section{Introduction}
The tunneling effect in a normal metal/superconductor 
(N/S) junction has been considered to be a basic quantum phenomenon 
since the discovery of superconductivity. 
Blonder, Tinkham, and Klapwijk (BTK) established that tunneling conductance 
can be expressed by the coefficients of the Andreev reflection 
and normal reflection in ballistic junctions
\cite{BTK82}. 
By extending the BTK theory, a conductance 
formula has been developed for unconventional 
superconductors \cite{TK95,TK96a}, where 
a pair potential changes sign on the Fermi surface and possesses 
the so-called surface Andreev-bound states (SABSs). 
This formula has clarified that the sharp 
zero-bias conductance peak (ZBCP) observed in many experiments of 
high $T_{C}$ cuprate \cite{Experiment1,Experiment2,Experiment3,
Experiment4,Experiment5,Experiment6,Experiment7}
stems from the zero energy surface Andreev-bound states (ZESABSs) 
\cite{ABS,ABSb,Hu,ABSR1,ABSR2} in unconventional nodal superconductors. 
Applying this formula for a spin-triplet 
chiral $p$-wave superconductor, a broad ZBCP has been obtained reflecting 
on the linear dispersion of the SABS \cite{YTK97,YTK98,Yakovenko}. 
This result is consistent with 
the tunneling experiments of Sr$_{2}$RuO$_{4}$ \cite{Mao,Kashiwaya11} and 
supports 
the realization of spin-triplet superconductivity in Sr$_{2}$RuO$_{4}$ 
\cite{Maeno,Maeno2}.  
\par
It is known that the physical origin of these SABSs 
stems from the chiral edge state protected by 
the topological invariant defined in the bulk Hamiltonian 
\cite{STYY11,Volovik97,FMS01}, 
and the high $T_{C}$ cuprate and Sr$_{2}$RuO$_{4}$ are regarded as 
topological superconductors \cite{tanaka12}. 
In the last decade, it has been established that 
topologically protected SABS 
can be generated based on a low-dimensional electron system with strong spin-orbit coupling 
without using unconventional pairings. 
For example, for the $s$-wave superconductor/ferromagnet junction 
on the surface of a topological insulator (TI), 
a chiral edge mode is generated similar to Sr$_{2}$RuO$_{4}$. 
Previously, one of the authors of this study, YT, derived a conductance formula 
for this hybrid system and clarified that the 
slope of the dispersion of the chiral edge mode is tunable by the gate voltage 
applied on the TI \cite{TYN09}. 
The derivation of the conductance formula is useful to capture the 
low energy charge transport in newly developed 
designed topological superconductors and superconducting TIs \cite{takami2014,Reeg}.

\begin{figure}[h]
    \centering
    \includegraphics[width=0.9\linewidth]{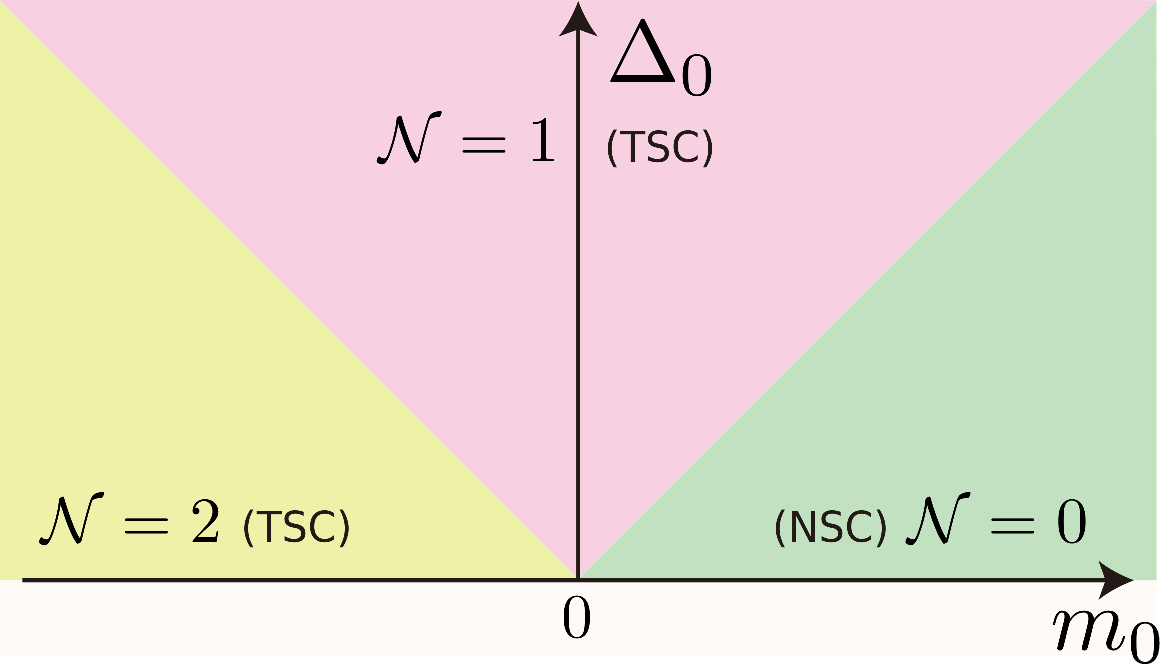}
    \caption{
        Phase diagram of BDI topological superconductor \cite{QHZ10,James}
        }
    \label{fg:phase_diagram}
\end{figure}

The AIII-class topological insulator has a winding number in one dimension, as shown in the 
topological periodic table \cite{Schnyder08}. 
By inducing the $s$-wave 
pair potential on an AIII-class TI, 
this system becomes a topological superconductor 
belonging to the BDI-class, which is characterized by the 
topological number  $\mathcal{N}$ \cite{James}. 
The phase diagram of this BDI topological superconductor is shown in 
Fig. \ref{fg:phase_diagram}. 
The present BDI topological superconductor
can be regarded as a one-dimensional 
version of the quantum anomalous Hall/superconductor hybrid system 
\cite{QHZ10}. 
Since there are many researches
about Quantum anomalous Hall / superconductor hybrid systems
\cite{Ji,Huang,He2017,Ii2011,Ii2012}, to clarify the BDI superconductor has a sufficient value.
It is remarkable that topological phase transition is tunable by changing the so-called mass parameter defined in the AIII topological insulator model, 
where a topological insulator is realized for $m<0$ \cite{QHZ10}. 
The number of edge modes, SABSs, in this BDI-class topological 
superconductor coincides with $\mathcal{N}$ \cite{QHZ10,James}. 
The zero-temperature conductance at
zero voltage shows a noteworthy feature. 
For the $\mathcal{N}=1$ state, 
the SABS appears as the single mode of the Majorana fermion; thus, the
resulting charge conductance becomes $2e^{2}/h$ \cite{James,YamakagePhysicaE}. 
Meanwhile, 
for $\mathcal{N}=2$, although the ZESABS exists as two Majorana fermions, 
because of the destructive interference between two Majorana fermions, 
the zero-bias conductance becomes zero \cite{James,YamakagePhysicaE}. 
Simultaneously, the reflection coefficients of the Andreev reflection 
disappear. 
Although several theoretical studies have been done regarding this 
system \cite{James,YamakagePhysicaE}, 
they are based on a low-energy 
effective model or numerical calculations in a finite system, 
and the charge conductance formula has not been derived analytically. 
It is beneficial to solve the scattering problem 
of a normal metal/one-dimensional (1D) BDI superconductor junction 
analytically and derive a conductance formula 
similar to other topological superconductors \cite{TYN09,takami2014}. \par

The aim of this study was to solve the Bogoliubov-de Gennes (BdG) equation 
of a normal metal/BDI superconductor junction based on the AIII TI in one dimension. 
We selected a standard normal metal with parabolic dispersion. 
We analytically obtained both the 
normal reflection coefficient $R_{\sigma\sigma'}$ and 
Andreev reflection coefficient $A_{\sigma\sigma'}$. 
Here, 
$\sigma$ is a spin index of a reflected electron (hole) for a normal (Andreev) 
reflection, and $\sigma'$ is a spin index of an injected electron. 
The resulting conductance shows a wide variety of line shapes. 
For the $\mathcal{N}=0$ state, the conductance exhibits a 
gap-like structure without sharp coherence peaks in contrast to the standard 
$U$-shaped line shape of differential conductance in tunneling spectroscopy of an $s$-wave superconductor. 
For the $\mathcal{N}=1$ state, $\sum_{\sigma\sigma'} R_{\sigma\sigma'} 
= \sum_{\sigma\sigma'} A_{\sigma\sigma'}$ 
is satisfied 
and the charge conductance has 
a quantized ZBCP of peak height $2e^{2}/h$. 
The width of this peak depends on the magnitude of $m_0$ and $\Delta_0$. 
For the $\mathcal{N}=2$ state, 
the charge conductance has a ZBCP splitting at $eV=0$ and 
becomes zero; this is consistent with previous results. 
Further, we clarified the spin rotation at zero bias voltage $eV=0$, 
when the quantization axis of the spin is along the $z$-axis.
For the $\mathcal{N}=1$ state, 
the spin direction of the normal-reflected electron and Andreev-reflected 
hole is directed along the $y$-axis. 
Meanwhile, for the $\mathcal{N}=2$ state, the Andreev reflection is 
absent and the spin rotation angle can become \degree{180} for the normal-reflected electron. 

The remainder of this paper is organized as follows. 
In section 2, we present the theory and the method to 
derive the conductance formula. 
In section 3, we detail the resulting 
conductance. 
In section 4, we demonstrate the spin rotation from the obtained reflection coefficients. Our results are summarized in section 5.

\section{Formulation}
We consider a normal metal/BDI superconductor (N/BDI) junction, as shown in Fig. 2 \cite{James}. 
\begin{figure}[h]
    \raggedright
    \includegraphics[width=0.9\linewidth]{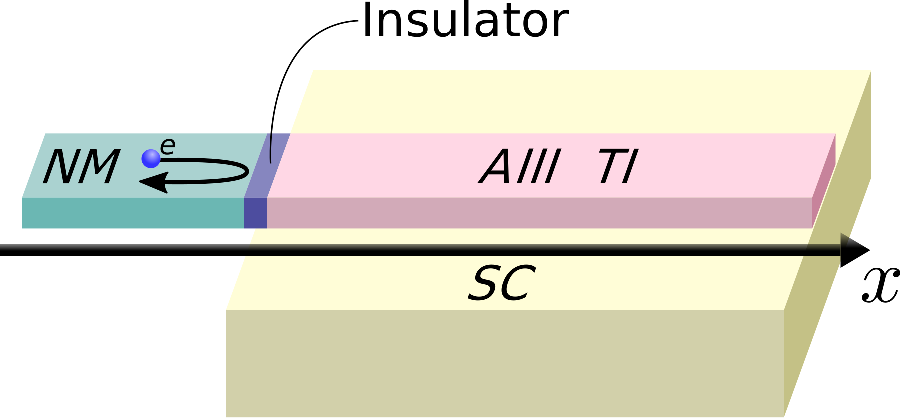}
    \caption{
        Normal metal/BDI superconductor junction. BDI superconductor is realized in the AIII topological insulator region that is proximity coupled with a spin-singlet $s$-wave superconductor \cite{James}.
        }
\end{figure}
The one-dimensional limit of a quantum anomalous Hall/superconductor hybrid system can be regarded as a BDI superconductor.
The Hamiltonian is given by 
\begin{align}
    H = H_{\rm N} \theta(-x) + U\delta(x) + H_{\rm SC}\theta(x), 
\end{align}
where $\theta(x)$ and $\delta(x)$ are the Heaviside step function and delta function, respectively. The second term indicates the barrier potential with 
barrier parameter $U$.

The Hamiltonian of a normal metal is defined as a standard free electron model with parabolic dispersion.
\begin{align}
    H_{\rm N}(k) = \left( \frac{\hbar^2}{2m_{\rm N}}k^2-\mu \right) \tau_z, 
\end{align}
where $\mu$, $m_{\rm N}$, and $\tau_z$ are 
the chemical potential, effective mass of the normal metal, 
and Pauli matrices in the Nambu space. 
Subsequently, the Fermi wave number is given by 
$k_{\rm F} = \sqrt{2m_{\rm N}\mu/\hbar^2}$. 
The Hamiltonian of the BDI superconductor is
\begin{align}
    H_{\rm SC}(k)=&
    \begin{pmatrix}
        h_{\rm AIII}(k) && i\sigma_y\Delta_0 \\
        -i\sigma_y\Delta_0 && -h_{\rm AIII}^*(-k) \\
    \end{pmatrix}
    ,\\
    h_{\rm AIII}(\bm{k}) =& m(k)\sigma_z + A_0k\sigma_x,\quad
    m(k) = m_0+B_0k^2, 
\end{align}
where $h_{\rm AIII}(k)$ is the AIII-class TI. 
$A_0$ and $B_0$ are material parameters.
$\Delta_0$ is the pair potential of the spin-singlet $s$-wave superconductor. 
Here, $m_{0}$, $\sigma_z$, and $A_0$ 
denote the effective mass, $z$ component of the Pauli matrix, and 
spin-orbital coupling, respectively. 
Herein, the chemical potential of the AIII-class insulator is fixed at zero. 
Therefore, the Fermi level is located in the middle between the conduction and valence band. For $m_0<0$, the AIII-class insulator becomes a topological hosting edge state 
\cite{QHZ10,James}.

We normalize each parameter using $k_{\rm F}$ in the remainder of this paper
\begin{align}
    \begin{cases}
        \tilde{m}_0 &= m_0/\mu \\
        \tilde{\Delta}_0 &= \Delta_0/\mu\\
    \end{cases}\quad
    \begin{cases}
        A &= A_0/\left(\frac{\hbar^2k_{\rm F}}{2m_N}\right)\\
        B &= B_0/\left(\frac{\hbar^2}{2m_N}\right)\\
        Z &= 2U/\left(\frac{\hbar^2k_{\rm F}}{2m_N}\right) 
    \end{cases}. 
\end{align}
The eigenenergy $E_{\pm}$ and 
eigenfunction $\psi_{\pm}(k)$ of $H_{\rm SC}$ are obtained:
\begin{align}
    E_{\pm} &= \sqrt{A_0^2k^2 + (m(k)\pm\Delta_0)^2}\label{eq:eigen_energy}\\
    \psi_{\pm}(k) &=
    \begin{pmatrix}
        1 \\ Q_{\pm}(k) \\ \pm Q_{\pm}(k) \\ \pm 1
    \end{pmatrix}
    ,\quad
    Q_{\pm}(k) = -\frac{m(k)\pm\Delta_0 - E}{A_0k}.
    \label{eq:wave_function}
\end{align}
We can confirm that the bulk energy gap of the BDI superconductor
closes at $m_0=\pm\Delta_0$, as calculated from the eigenenergy. 
Thus, the topological phase transition occurs at $m_0=\pm\Delta_0$. 

In the following section, we study the scattering problem of the N/BDI junction. 
We assume that the spin direction of an injected electron is 
along the $z$ axis. 
In the 1D BDI superconductor, the wave function is satisfied, as follows: 
\begin{align}
    \Psi_{\rm N}(x) &= \Psi_{\rm in}(x) + \Psi_{\rm ref}(x) & (x<0)\\
    \Psi_{\rm SC}(x) &= \Psi_{\rm tra}(x) & (x>0)
\end{align}
\begin{align}
    \Psi_{\rm in} &= \begin{pmatrix}\delta_{\uparrow\sigma}\\\delta_{\downarrow\sigma}\\0\\0\end{pmatrix}e^{ik_{\rm F}x},\quad
    \Psi_{\rm ref} = \begin{pmatrix}b_{\uparrow\sigma}\\b_{\downarrow\sigma}\\0\\0\end{pmatrix}e^{-ik_{\rm F}x} + \begin{pmatrix}0\\0\\a_{\uparrow\sigma}\\a_{\downarrow\sigma}\end{pmatrix}e^{ik_{\rm F}x},\quad\\
    \Psi_{\rm tra} =& t_{1+}\psi_{\pm}(k_{1+})e^{ik_{1+}x}
    + t_{2+}\psi_{\pm}(k_{2+})e^{ik_{2+}x}\nonumber\\
    &+ t_{1-}\psi_{\pm}(k_{1-})e^{ik_{1-}x}
    + t_{2-}\psi_{\pm}(k_{2-})e^{ik_{2-}x}, 
\end{align}
where $\sigma$ is the spin of an injected electron; 
$b_{\sigma'\sigma}$ and $a_{\sigma'\sigma}$ are the amplitudes of 
the normal and Andreev reflections with $\sigma'=\uparrow(\downarrow)$; 
$t_{1\pm}$ and $t_{2\pm}$ are the corresponding transmission amplitudes. 
The wave number in the BDI superconductor is calculated from the eigenenergy: 
$E_{\pm}$

\begin{align}\label{eq:wave_number1}
    \left(k_{1,\pm}(E) \right)^2 =& \frac{1}{2B_0^2}\left(-(2B_0(m_0\pm\Delta_0)+A_0^2)\right.\nonumber\\
    &\left.+\sqrt{A_0^4+4B_0(m_0\pm\Delta_0)A_0^2+4B_0^2 E^2}\right)\\
    \label{eq:wave_number2}
    \left(k_{2,\pm}(E)\right)^2 =& \frac{1}{2B_0^2}\left(-(2B_0(m_0\pm\Delta_0)+A_0^2)\right.\nonumber\\
    &\left.-\sqrt{A_0^4+4B_0(m_0\pm\Delta_0)A_0^2+4B_0^2E^2}\right),
\end{align}
where $E$ is the energy measured from the Fermi level. 
The sign of the wave number is determined by the group velocity 
such that the wave function does not diverge for $x \rightarrow \infty$. 
The boundary condition of the wave function at $x=0$ is given as follows: 
\begin{align}
    \begin{cases}
        \Psi_{\rm SC}(x=0) - \Psi_{\rm N}(x=0) = 0\\
        \hbar(\hat{v}_{\rm SC}\left\{\Psi_{\rm SC}(x)\right\}|_{x=+0}  -
        \hat{v}_{\rm N}\left\{\Psi_{\rm N}(x)\right\}|_{x=-0}) \\
        \hfill= -2iU\tau_z\Psi_{\rm N}(x=0)
    \end{cases}.
    \label{eq:bound_cond}
\end{align}
Here, $\hat{v}$ is 
the velocity operator 
$\hat{v}=\frac{1}{\hbar}\frac{\partial H}{\partial (-i\partial_x)}$.

\section{Tunneling Effect}
The charge conductance $\Gamma$ in the N/BDI junction can be expressed using the reflection coefficients 
\begin{align}
    \Gamma = \frac{e^2}{h}\left(2 - \sum_{\sigma,\sigma'} \left(R_{\sigma\sigma'} - A_{\sigma\sigma'}\right)\right). 
\end{align}
The amplitude of the normal reflection $\bm{b}_\sigma=(b_{\uparrow\sigma},b_{\downarrow\sigma})$
and that of the Andreev reflection 
$\bm{a}_\sigma=(a_{\uparrow\sigma},a_{\downarrow\sigma})$ 
for an injected electron with $\sigma=\uparrow,\downarrow$ 
are expressed by 
two component vectors obtained from the 
boundary condition \eqref{eq:bound_cond}
\begin{align}
    \bm{b}_{\sigma} =& \left(\bm{I}+iZ\left\{\left(\hat{\bm{K}}_{+}+\gamma^{*}\bm{I}\right)^{-1}+\left(\hat{\bm{K}}_{-}+\gamma^{*}\bm{I}\right)^{-1}\right\}\right)^{-1}\nonumber\\
    &\left(\bm{I}-\gamma^*\left\{\left(\hat{\bm{K}}_{+}+\gamma^{*}\bm{I}\right)^{-1}+\left(\hat{\bm{K}}_{-}+\gamma^{*}\bm{I}\right)^{-1}\right\}\right)\bm{u}_{\sigma}\\
    \bm{a}_{\sigma} =& -2\sigma_x\left(\bm{I}+iZ\left\{\left(\hat{\bm{K}}_{+}+\gamma\bm{I}\right)^{-1}+\left(\hat{\bm{K}}_{-}+\gamma\bm{I}\right)^{-1}\right\}\right)^{-1}\nonumber\\
    &\left(\left\{\left(\hat{\bm{K}}_{+}+\gamma\bm{I}\right)^{-1}-\left(\hat{\bm{K}}_{-}+\gamma\bm{I}\right)^{-1}\right\}\right)\bm{u}_{\sigma}, 
\end{align}
with $\bm{u}_{\sigma}=(\delta_{\uparrow\sigma},\delta_{\downarrow\sigma})$. 
Here, $\gamma\equiv2+iZ$ with the barrier parameter $Z$,
$\bm{I}$ is a $2\times2$ unit matrix,  
and $\hat{\bm{K}}_{\pm}$ is a $2\times2$ matrix 
with respect to the wave number \eqref{eq:wave_number1}\eqref{eq:wave_number2} 
using the factor of the wave function \eqref{eq:wave_function}.
\begin{align}
\hat{\bm{K}}_{\pm} \equiv& A\sigma_x + 2B\sigma_z\hat{\bm{Q}}_{\pm}\begin{pmatrix}k_{1\pm}/k_{\rm F} && 0 \\ 0 && k_{2\pm}/k_{\rm F} \\ \end{pmatrix}\hat{\bm{Q}}_{\pm}^{-1}\\
\hat{\bm{Q}}_{\pm} \equiv& \begin{pmatrix} 1 && 1 \\ Q_{\pm}(k_{1\pm}) && Q_{\pm}(k_{2\pm})\end{pmatrix}. 
\end{align}
The matrices of the reflection coefficients 
of the normal reflection $R_{\sigma\sigma'}$ and 
that of the Andreev reflection  $A_{\sigma\sigma'}$
are given by 
\begin{equation}
R_{\sigma\sigma'}
=
\begin{pmatrix}
\mid b_{\uparrow \uparrow} \mid^{2} & \mid b_{\uparrow \downarrow} \mid^{2} \\
\mid b_{\downarrow \uparrow} \mid^{2} & \mid b_{\downarrow \downarrow} \mid^{2}
\end{pmatrix}
\end{equation}
\begin{equation}
A_{\sigma\sigma'}
=
\begin{pmatrix}
\mid a_{\uparrow \uparrow} \mid^{2} & \mid a_{\uparrow \downarrow} \mid^{2} \\
\mid a_{\downarrow \uparrow} \mid^{2} & \mid a_{\downarrow \downarrow} \mid^{2}
\end{pmatrix},
\end{equation}

\begin{figure}[htbp]
    \raggedright
    \includegraphics[width=\linewidth]{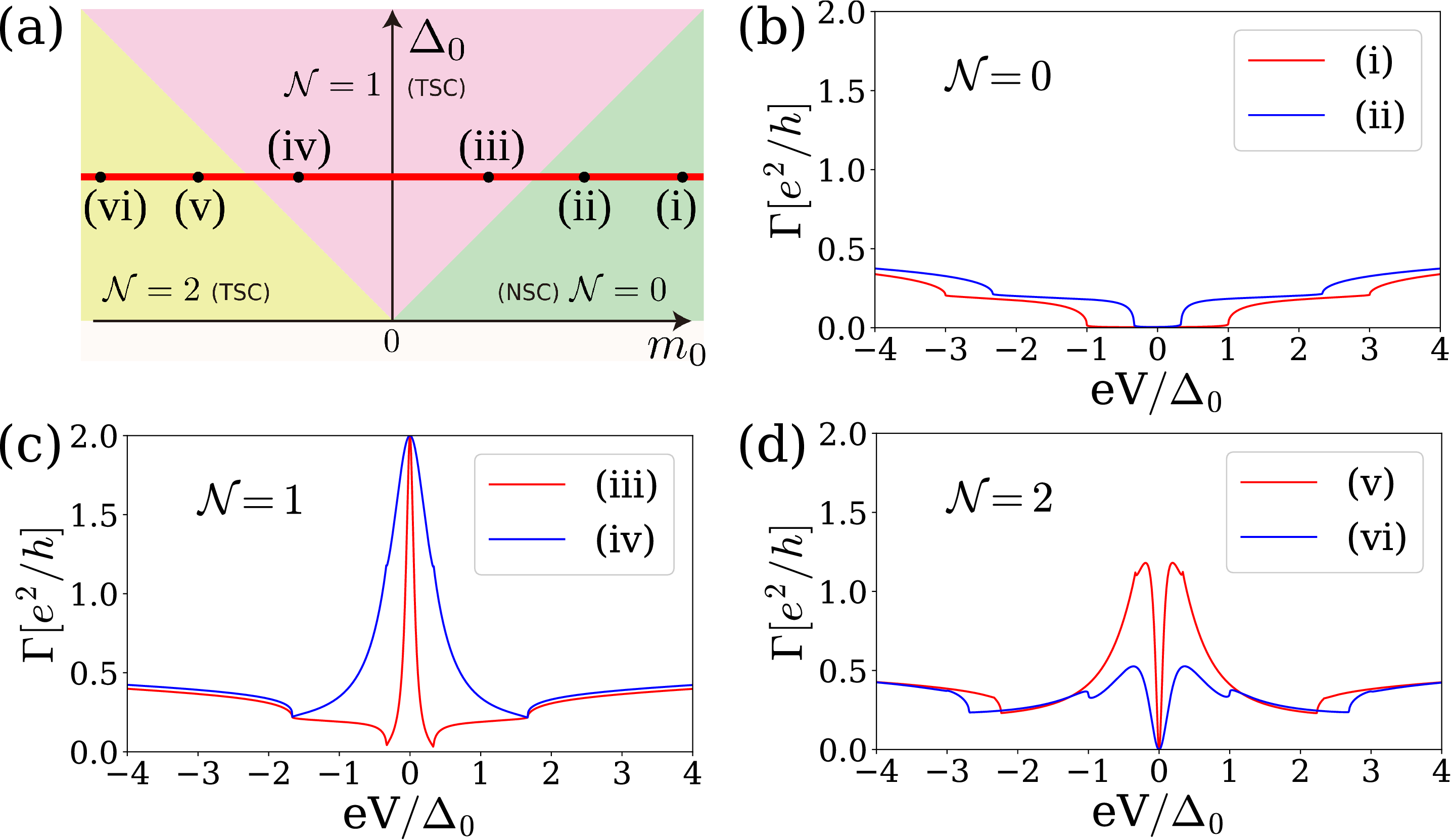}
    \caption{
        Calculated conductance for various $m_{0}$. 
        We selected $\tilde{\Delta}_0=0.03$, $A=B=0.1$, and $Z=1$. 
        (a): The values of $(m_0,\Delta_0)$ in the phase diagram of the BDI superconductor. 
        (b): $\mathcal{N}=0$ with  (i)$\tilde{m}_0=0.06$ and (i)$\tilde{m}_0=0.04$. 
        (c): $\mathcal{N}=1$ with  (iii)$\tilde{m}_0=0.02$ and (iv)$\tilde{m}_0=-0.02$.
        (d): $\mathcal{N}=2$ with  (v)$\tilde{m}_0=-0.04$ and (vi)$\tilde{m}_0=-0.06$.
    }
    \label{fg:1D_cond}
\end{figure}

We calculate the conductance $\Gamma$ analytically 
with bias voltage $V$ where $E=eV$ is satisfied. 
We selected various mass parameters $m_{0}$ for a fixed $\Delta_0$, 
as shown in Fig.\ref{fg:1D_cond}. 
For $\mathcal{N}=0$ cases, the obtained conductance never becomes exactly zero for any $V$ ($\Gamma\not{=}0$) 
owing to the Andreev reflection. 
The sharp coherent peaks that appear in the case of conventional tunneling spectroscopy of the $s$-wave superconductor is absent in the conductance. 
This is because the spin-singlet $s$-wave pair potential is induced in the 
insulating AIII phase. 
The conductance shows a two-gap behavior at $eV=|m_0\pm\Delta_0|$.
For the $\mathcal{N}=1$ state, the ZBCP appears with its peak height of $2e^2/h$.
The peak width is determined by the barrier parameter $Z$ and gap width $W_{-}$, which is the gap width of the energy band $E_{-}$ defined in eq. \eqref{eq:eigen_energy}
\begin{align}
    W_{-} = 
    \begin{cases}
        \left|m_0-\Delta_0\right| & \left( m_0-\Delta_0>-A_0^2/2B_0\right)\\
        \sqrt{\frac{A_0^2}{B_0}\left(\left|m_0-\Delta_0\right|-\frac{A_0^2}{4B_0}\right)} & \left(m_0-\Delta_0<-A_0^2/2B_0\right)\\
    \end{cases}. 
\end{align}
When $m_{0}$ decreases, $W_{-}$ increases in $\mathcal{N}=1$ phase, and peak width, which is approximately proportional to $W_{-}$, also increases (Fig.3(c)). 
At $m_0=\Delta_0$, the width of the peak becomes zero and the 
energy spectrum of the BDI superconductor becomes gapless corresponding to 
a topological transition. 
This ZBCP is due to the ZESABS that manifests as 
a Majorana fermion at the edge of the BDI superconductor \cite{QHZ10,James}. 
For the $\mathcal{N}=2$ state, 
the charge conductance exhibits a ZBCP splitting. 
The height of the peaks at a nonzero voltage is suppressed with the decrease in the 
value of $m_{0}$.  
At $eV=0$, the conductance at zero voltage becomes exactly zero, 
consistent with previous results \cite{Ii2012,James,YamakagePhysicaE}.

\section*{Zero Bias Voltage}
Next, we focus on the conductance at zero voltage where more 
compact formula of coefficients of the Andreev and normal reflections 
and conductance can be available. 
For convenience, we introduce the following:
\begin{align}
    D_{\pm} &\equiv \sqrt{A^2 + 4B(\tilde{m}_0\pm\tilde{\Delta}_0)}.
\end{align}
The amplitudes of both the normal and Andreev reflections can be expressed as 
\begin{alignat}{2}
    &\begin{cases}
        \bm{b}_{\sigma}=&\frac{(\gamma^{*2}+D_+D_-)}{(|\gamma|^2-D_+D_-)^2+4(D_++D_-)^2}\nonumber\\
        &\left\{(|\gamma|^2-D_+D_-) - 2i(D_++D_-)\sigma_z\right\}\bm{u}_{\sigma}\\
        \bm{a}_{\sigma}=&-\frac{2(D_+-D_-)}{(|\gamma|^2-D_+D_-)^2+4(D_++D_-)^2}\nonumber\\
        &\left\{2(D_++D_-)\sigma_x+(|\gamma|^2-D_+D_-)\sigma_y \right\}\bm{u}_{\sigma}\\
    \end{cases}&(\text{if }\mathcal{N}=0) \\
    &\begin{cases}
        \bm{b}_{\sigma}=&\frac{1}{2}\left\{\frac{\gamma^{*2}-D_+^2}{|\gamma|^2+D_+^2}-{\rm sgn}(A_0)\frac{\gamma^{*2}+D_+^2}{|\gamma|^2+D_+^2}\sigma_y\right.\nonumber\\
        &\left.-i\frac{2\gamma^*D_+}{|\gamma|^2+D_+^2}\sigma_z\right\}\bm{u}_{\sigma}\\
        \bm{a}_{\sigma}=&-\frac{i}{2}\left\{{\rm sgn}(A_0)\frac{2ZD_+}{|\gamma|^2+D_+^2}+i\frac{|\gamma|^2-D_+^2}{|\gamma|^2+D_+^2}\sigma_x\right.\nonumber\\
        &\left.-i\frac{4D_+}{|\gamma|^2+D_+^2}\sigma_y-{\rm sgn}(A_0)\sigma_z\right\}\bm{u}_\sigma
    \end{cases}&(\text{if }\mathcal{N}=1) \\
    &\begin{cases}
        \bm{b}_{\sigma}=&-\frac{1}{\gamma}\left(iZ+2{\rm sgn}(A_0)\sigma_y\right)\bm{u}_\sigma\\
        \bm{a}_{\sigma}=&\bm{0}
    \end{cases}&(\text{if }\mathcal{N}=2). 
    \label{eq:ref_amp}
\end{alignat}
For $\mathcal{N}=0$, both amplitudes of 
the normal and Andreev reflections exist. 
For $\mathcal{N}=1$, 
after some straightforward calculations, we obtain 
\begin{align}
    \sum_{\sigma,\sigma'}R_{\sigma\sigma'} = \sum_{\sigma,\sigma'}A_{\sigma\sigma'}.
\end{align}
This implies that the contributions of the normal and 
Andreev reflections are completely balanced. 
This property is unique although the ZBCP exists 
in both the Andreev and normal reflections. 
It is qualitatively different from the previous normal metal/unconventional 
junctions with the perfect resonant case, where 
only the Andreev reflection exists at zero voltage 
\cite{TK95,ABSR1}. 
Consequently, differences are observed between the heights of these ZBCPs. 
For the $\mathcal{N}=2$ state, 
although the ZESABS exists, the Andreev reflection is completely suppressed. 

\begin{figure}[!htbp]
    \raggedright
    \includegraphics[width=\linewidth]{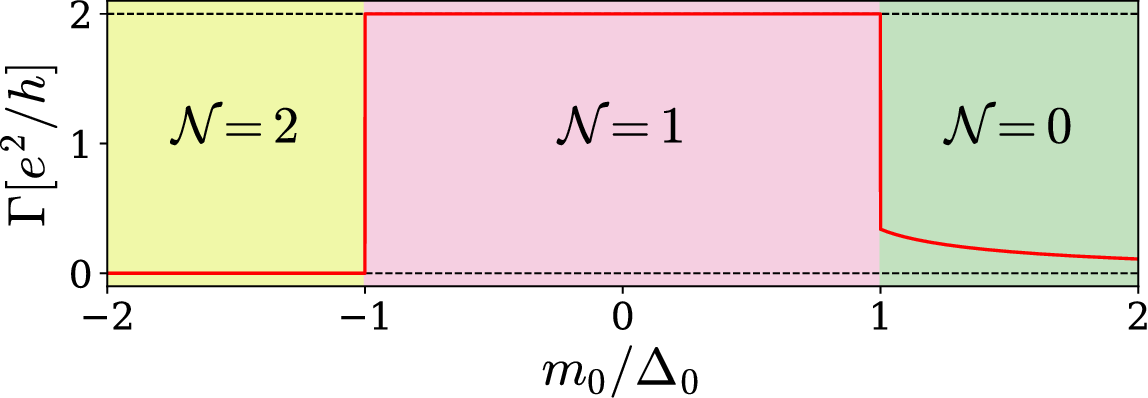}
    \caption{
        Conductance with zero bias voltage for $\tilde{\Delta}_0=0.25$, $A=B=0.1$, $Z=1$ (Solid line). Dashed lines are the guide for the eye.
        }
    \label{fg:cond_E0}
\end{figure}

Using the results above, we obtain $\Gamma$ as follows:
\begin{align}
    \Gamma = \frac{e^2}{h}\times
    \begin{cases}
        \frac{16(D_+-D_-)^2}{(|\gamma|^2-D_+D_-)^2+4(D_++D_-)^2} & (\text{if }\mathcal{N}=0)\\
        2 &(\text{if }\mathcal{N}=1)\\
        0 &(\text{if }\mathcal{N}=2)
    \end{cases}. 
\end{align}
For the $\mathcal{N}=1$ state, we can demonstrate analytically that 
the charge conductance becomes $\frac{2e^2}{h}$, as shown in Fig.\ref{fg:cond_E0}.

\section{Spin Rotation}
We have obtained the normal and Andreev reflection coefficients analytically; thus, we can analyze the detailed property of the reflected particles. 
In this section, we study the spin rotation through the scattering processes at the interface. Here, we consider the normal and Andreev reflections at zero bias voltage to use the formulae in the previous section.
Subsequently, the reflection amplitude $\bm{b}_\sigma$, $\bm{a}_\sigma$ is expressed by the spin rotational operator $\exp\left(i\frac{\hat{\theta}}{2}\cdot\hat{\sigma}\right)$, where $\hat{\theta}=\theta\hat{\bm{n}}$ denotes the rotational axis $\hat{\bm{n}}$ and rotational angle $\theta$. The results depend highly on the topological phase in the BDI superconductor.

For the $\mathcal{N}=1$ state, the reflection coefficients are denoted by a linear combination of spinors that are expressed by two types of spin rotations
\begin{align}
    \begin{cases}
        \bm{b}_{\sigma}=\frac{1}{2}\left\{\exp\left(i\frac{\hat{\theta}_{1b1}}{2}\cdot\hat{\sigma}\right)-i\exp\left(i\frac{\hat{\theta}_{1b2}}{2}\cdot\hat{\sigma}\right)\right\}\bm{u}_{\sigma}\\
        \bm{a}_{\sigma}=\frac{{\rm sgn}(A)}{2}\left\{\exp\left(i\frac{\pi}{2}\sigma_z\right)-i\exp\left(i\frac{\hat{\theta}_{1a}}{2}\cdot\hat{\sigma}\right)\right\}\bm{u}_\sigma
    \end{cases}&(\text{if }\mathcal{N}=1),
    \label{eq:spin_N=1}
\end{align}
where the rotation angles are as follows:

\begin{align}
    \hat{\theta}_{1b1}/2=&\arctan\left(\frac{4\sqrt{Z^2+D_{+}^2}}{4-Z^2-D_{+}^2}\right)\begin{pmatrix}0\\{\rm sgn}(A)Z\\-D_{+}\end{pmatrix}\Bigg/ \left|\begin{pmatrix}0\\{\rm sgn}(A)Z\\-D_{+}\end{pmatrix}\right|\\
    \hat{\theta}_{1b2}/2=&\arctan\left(-\frac{\sqrt{(4-Z^2+D_{+}^2)^2+4Z^2D_{+}^2}}{4Z}\right)\nonumber\\&\begin{pmatrix}0\\(4-Z^2+D_{+}^2){\rm sgn}(A)\\2ZD_{+}\end{pmatrix}\Bigg/\left|\begin{pmatrix}0\\(4-Z^2+D_{+}^2){\rm sgn}(A)\\2ZD_{+}\end{pmatrix}\right| \\
    \hat{\theta}_{1a}/2=&\arctan\left({\rm sgn}(A)\frac{\sqrt{|\gamma|^4+2D_{+}^2(4-Z^2)+D_{+}^4}}{2ZD_{+}}\right)\nonumber\\&\begin{pmatrix}|\gamma|^2-D_{+}^2\\-4D_{+}\\0\end{pmatrix}\Bigg/\left|\begin{pmatrix}|\gamma|^2-D_{+}^2\\-4D_{+}\\0\end{pmatrix}\right|. 
\end{align}

\begin{figure}[!htbp]
    \raggedright
    \includegraphics[width=\linewidth]{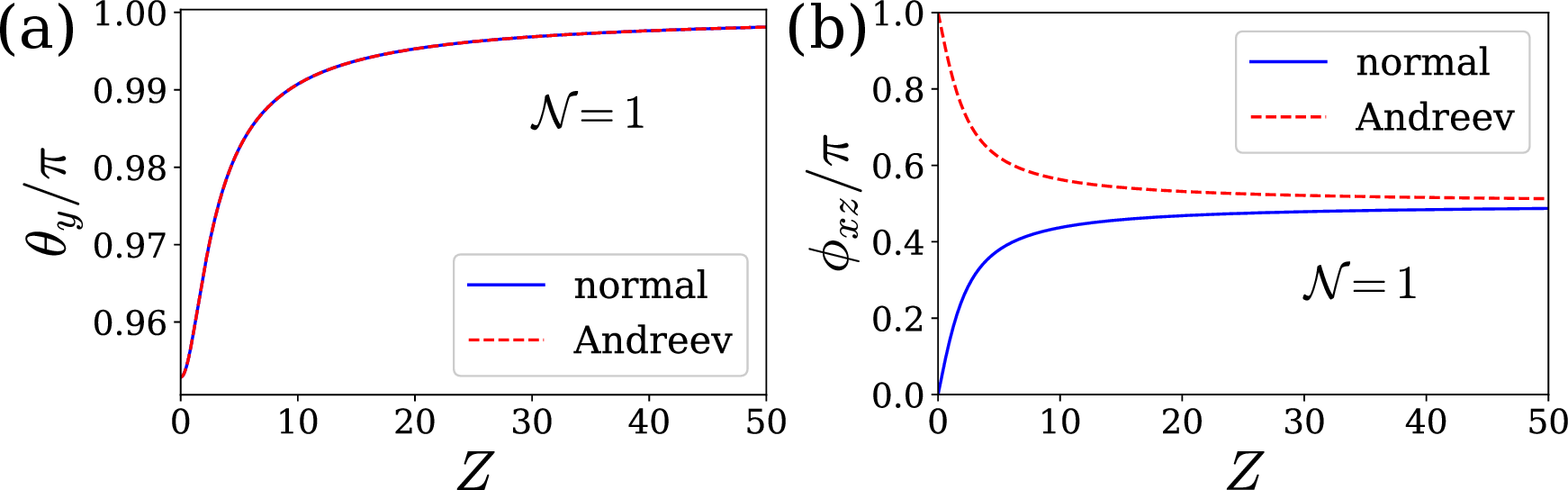}
    \caption{
        Calculated spin rotation for $\mathcal{N}=1$ at $eV=0$. 
        We selected $\tilde{\Delta}_0=0.03$, $A=B=0.1$, $\tilde{m}_0=0$ and $Z=1$. 
        (a): Plot of $\theta_{y}$ that is the polar angle of the spin direction from the $y$-axis. 
        The normal and Andreev reflections has the same $\theta_{y}$ value.
        (b): Plot of $\phi_{xz}$ that is the azimuth angle of the spin direction in the $xz$-plane.
    }
    \label{fg:spin_1}
\end{figure}

Fig.\ref{fg:spin_1} shows the spin direction of the normal and Andreev reflections 
with the injection of an up-spin electron, where $\theta_{y}$ and $\phi_{xz}$ are the polar angle from the $y$-axis and the azimuth angle in the $xz$-plane, respectively. 
In the case of $A>0$, the spin direction for both the normal and Andreev reflections are almost along the direction of $-\hat{\bm{y}}$ because of $\theta_{y}\sim\pi$ for any $Z$, as shown in Fig.\ref{fg:spin_1}(a). 
Here, $\hat{\bm{y}}$ is a unit vector along the $y$-direction. 
Meanwhile, the spin directions are the opposite in the $A<0$ case. 
This implies that the spin direction of the reflected waves depends on that of the BDI superconductor which couples to the momentum by the spin-orbit coupling.
Additionally, we confirm that the spin direction for the normal and Andreev reflections with a down-spin injection is the same as those for an up-spin injection. 
In other words, the spin directions of the reflected waves are polarized both in the electron and hole sectors. 
This spin polarization phenomenon is caused by the spin-orbit coupling of the BDI superconductor.

\begin{figure}[!htbp]
    \raggedright
    \includegraphics[width=\linewidth]{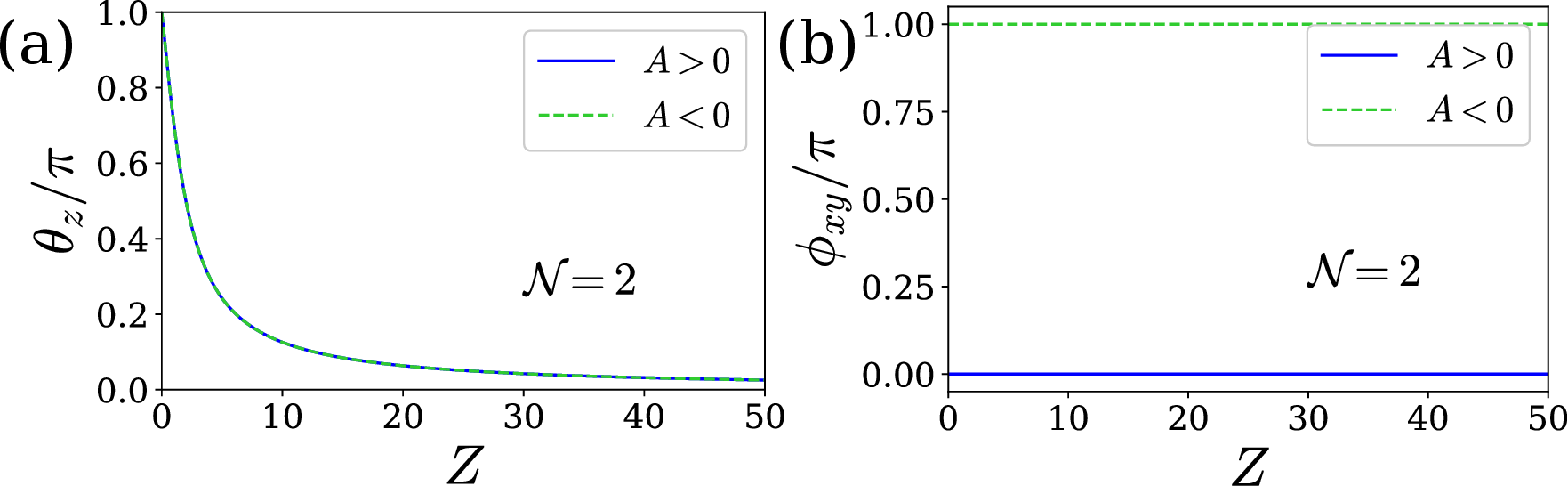}
    \caption{
        Calculated spin rotation injected up-spin for $\mathcal{N}=2$ at $eV=0$. 
        We selected $\tilde{\Delta}_0=0.03$, $A=B=0.1$, and $Z=1$. 
        (a): Plot of $\theta_{z}$ that is the polar angle of the spin direction from the $z$-axis. 
        In the $A>0$ and $A<0$ cases, the same value is observed for $\theta_{z}$.
        (b): Plot of $\phi_{xy}$ that is the azimuth angle of the spin direction in the $xy$-plane.
    }
    \label{fg:spin_2}
\end{figure}

For the $\mathcal{N}=2$ state, the reflection amplitudes are denoted as follows: 

\begin{align}
    &\begin{cases}
        \bm{b}_{\sigma}&=R_{2b}\exp\left(i\frac{\hat{\theta}_{2b}}{2}\cdot{\sigma}\right)\bm{u}_\sigma\\
        \bm{a}_{\sigma}&=\bm{0}
    \end{cases}&(\text{if }\mathcal{N}=2), 
\end{align}
where the coefficient and rotation angle are as follows:
\begin{align}
    &R_{2b} = -i\frac{2-iZ}{\sqrt{4+Z^2}},\quad\hat{\theta}_{2b}/2=\arctan\left(-{\rm sgn}(A)\frac{2}{Z}\right)\begin{pmatrix}0\\1\\0\end{pmatrix}. 
    \label{eq:spin_N=2} 
\end{align}
According to Eq.(\ref{eq:spin_N=2}), normal reflection depends only on the sign of $A$ and does not depend on other parameters of the BDI superconductor. 
Fig.\ref{fg:spin_2} shows the spin direction of the normal reflection in an up-spin injection, where $\theta_{z}$ and $\phi_{xy}$ are the polar angle from the $z$-axis and the azimuth angle in the $xy$-plane, respectively. 
As shown in Fig.\ref{fg:spin_2}(a), the spins of the reflections are directed to $-\hat{\bm{z}}$, i.e., $\pi$-rotation at $Z=0$, but do not change through the scattering for  $Z=\infty$. 
Here, $\hat{\bm{z}}$ is a unit vector. 
This is because the couplings between the injected electron and the edge state of the BDI becomes weak with increasing barrier strength.
From $Z=0$ to $\infty$, the spin direction of the normal reflection rotates in the $xz$-plane. It is noteworthy that the azimuth angle depends on the sign of $A$, as shown in Fig.\ref{fg:spin_2}(b). 
It is known that a giant spin rotation appears in the normal metal/quantum spin Hall junction depending on the edge states\cite{YTN09}. 
Similarly, spin rotations appear for a weak barrier strength. 

Finally, for the $\mathcal{N}=0$ state, the reflection amplitudes are expressed as follows:

\begin{align}
    &\begin{cases}
        \bm{b}_{\sigma}&=R_{0b}\exp\left(i\frac{\hat{\theta_{0b}}}{2}\cdot\hat{\sigma}\right)\bm{u}_{\sigma}\\
        \bm{a}_{\sigma}&=R_{0a}\exp\left(i\frac{\hat{\theta}_{0a}}{2}\cdot\hat{\sigma} \right)\bm{u}_{\sigma}\\
    \end{cases}&(\text{if }\mathcal{N}=0), 
\end{align}
where the coefficients and rotation angles are as follows:

\begin{align}
    &\begin{cases}
        R_{0b} =& \frac{\gamma^{*2}+D_{+}D_{-}}{\sqrt{(|\gamma|^2-D_{+}D_{-})^2+4(D_{+}+D_{-})^2}}\\
        \hat{\theta}_{0b}/2=&\arctan\left(\frac{2(D_{+}+D_{-})}{|\gamma|^2-D_{+}D_{-}}\right)\begin{pmatrix}0\\0\\1\end{pmatrix} \\
    \end{cases}\\
    &\begin{cases}
        R_{0a} =& \frac{2i(D_{+}-D_{-})}{\sqrt{(|\gamma|^2-D_{+}D_{-})^2+4(D_{+}+D_{-})^2}}\\
        \hat{\theta}_{0a}/2=&\frac{\pi}{2}\begin{pmatrix}2(D_{+}+D_{-})\\|\gamma|^2-D_{+}D_{-}\\0\end{pmatrix} \Bigg/ \left|\begin{pmatrix}2(D_{+}+D_{-})\\|\gamma|^2-D_{+}D_{-}\\0\end{pmatrix}\right|\\
    \end{cases}. 
\end{align}
When the spin direction for the injected electrons is along the $\hat{\bm{z}}$-axis, any spin flipping or rotation does not appear because $\theta_{0b}$ directs $\hat{\bm{z}}$ and the rotational axis is along the $z$-axis. 
Further, the Andreev reflection is flipped simply, i.e., $\uparrow$ to $\downarrow$ or vice versa. 
This spin rotation is similar to that of conventional tunneling of the $s$-wave superconductor. The results above are summarized in Table I.

\begin{table}[!htb]
    \label{}
    \caption{
        Spin direction of the normal and Andreev reflections with injected up-spin electron. $\hat{\bm{y}}$ and $\hat{\bm{z}}$ are unit vectors
        }
    \centering
    \begin{tabular}{|c||c|c|c|c|} \hline
        \multirow{2}{*}{$\mathcal{N}$} &\multicolumn{2}{|c|}{Normal} & \multicolumn{2}{|c|}{Andreev} \\ 
        &$Z=0$ & $Z=\infty$ & $Z=0$ & $Z=\infty$ \\ \hline
        $0$ & $+\hat{\bm{z}}$ & $+\hat{\bm{z}}$ & $-\hat{\bm{z}}$ & $-\hat{\bm{z}}$ \\ \hline
        $1$ & $-{\rm sgn}(A)\hat{\bm{y}}$ & $-{\rm sgn}(A) \hat{\bm{y}}$ & $-{\rm sgn}(A) \hat{\bm{y}}$ & $-{\rm sgn}(A) \hat{\bm{y}}$\\ \hline
        $2$ & $-\hat{\bm{z}}$ & $+\hat{\bm{z}}$ & - & - \\ \hline
    \end{tabular}
\end{table}

\section{Conclusion}
We have studied the tunneling effect in a topological superconductor 
based on a 1D AIII-class TI that is proximity 
coupled with a spin-singlet $s$-wave superconductor. 
This topological superconductor belongs to the BDI-class and 
topologically different phases are 
characterized by the topological invariant $\mathcal{N}$ for 
a bulk BDI superconductor. 
By solving the BdG equation 
of the normal metal/BDI superconductor junction, 
we have analytically obtained both the 
 normal reflection coefficient $R_{\sigma\sigma'}$ and 
Andreev reflection coefficient $A_{\sigma\sigma'}$, 
where $\sigma(\sigma')$ is a spin index of the reflected (injected) wave. 
The resulting conductance indicates a wide variety of line shapes. 
For the $\mathcal{N}=0$ state, the obtained conductance exhibits a 
Gap-like structure without sharp coherence peaks at their maxima, 
in contrast to the standard $U$-shaped line shape 
in $s$-wave superconductor tunneling spectroscopy. 
For the $\mathcal{N}=1$ state, $\sum_{\sigma\sigma'} R_{\sigma\sigma'} 
= \sum_{\sigma\sigma'} A_{\sigma\sigma'}$ 
is satisfied. 
The obtained conductance exhibits 
a ZBCP of height $2e^{2}/h$. 
The width of this peak depends on the magnitudes of $m_0$ and $\Delta_0$. 
With the decrease in the value of $m_0$ for a fixed $\Delta_0$, 
the width of the peak increases up to $m_0=-\Delta_0$. 
For the $\mathcal{N}=2$ state, 
the charge conductance exhibits a ZBCP spitting; at $eV=0$, 
it became zero, consistent with previous results. 
Further, we have calculated the spin rotation at zero bias voltage $eV=0$, 
when the quantization axis of the spin is along the $z$-axis.
For the $\mathcal{N}=0$ state, the spin direction of the reflected electron is 
along the $z$-axis and that of the hole is in the opposite direction. 
For the $\mathcal{N}=1$ state, 
the spin directions of the reflected electron and hole 
are directed along the $y$-axis. 
Meanwhile, for the $\mathcal{N}=2$ state, 
$A_{\sigma\sigma'}=0$ is always satisfied and 
the reflected electron exhibited a spin rotation. 
The spin rotation angle can become \degree{180} in the extreme case 
when no barrier exists at the boundary.

\section*{Acknowledgments}
We would like to thank valuable discussion with A. Yamakage. 
This work was supported by  Grant-in-Aid
for Scientific Research on Innovative Areas, Topological
Material Science (Grant Nos. JP15H05851, 
JP15H05853) and JSPS KAKENHI Grant Numbers JP18K03538 and JP18H01176
from the Ministry of Education, Culture,
Sports, Science, and Technology, Japan (MEXT).

\appendix
\bibliographystyle{jpsj}
\bibliography{paper.bbl}

\begin{thebibliography}{10}

\bibitem{BTK82}
G.~E. Blonder, M.~Tinkham, and T.~Klapwijk: Phys. Rev. B {\bfseries 25} (1982)
  4515.

\bibitem{TK95}
Y.~Tanaka and S.~Kashiwaya: Phys. Rev. Lett. {\bfseries 74} (1995) 3451.

\bibitem{TK96a}
Y.~Tanaka and S.~Kashiwaya: Phys. Rev. B {\bfseries 53} (1996) 9371.

\bibitem{Experiment1}
S.~Kashiwaya, Y.~Tanaka, M.~Koyanagi, H.~Takashima, and K.~Kajimura: Phys. Rev.
  B {\bfseries 51} (1995) 1350.

\bibitem{Experiment2}
S.~Kashiwaya, Y.~Tanaka, N.~Terada, M.~Koyanagi, S.~Ueno, L.~Alff,
  H.~Takashima, Y.~Tanuma, and K.~Kajimura: J. Phys. Chem. Solid {\bfseries 59}
  (1998) 2034.

\bibitem{Experiment3}
M.~Covington, M.~Aprili, E.~Paraoanu, L.~H. Greene, F.~Xu, J.~Zhu, and C.~A.
  Mirkin: Phys. Rev. Lett. {\bfseries 79} (1997) 277.

\bibitem{Experiment4}
L.~Alff, H.~Takashima, S.~Kashiwaya, N.~Terada, H.~Ihara, Y.~Tanaka,
  M.~Koyanagi, and K.~Kajimura: Phys. Rev. B {\bfseries 55} (1997) R14757.

\bibitem{Experiment5}
J.~Y.~T. Wei, N.-C. Yeh, D.~F. Garrigus, and M.~Strasik: Phys. Rev. Lett.
  {\bfseries 81} (1998) 2542.

\bibitem{Experiment6}
A.~Biswas, P.~Fournier, M.~M. Qazilbash, V.~N. Smolyaninova, H.~Balci, and
  R.~L. Greene: Phys. Rev. Lett. {\bfseries 88} (2002) 207004.

\bibitem{Experiment7}
B.~Chesca, H.~J.~H. Smilde, and H.~Hilgenkamp: Phys. Rev. B {\bfseries 77}
  (2008) 184510.

\bibitem{ABS}
L.~J. Buchholtz and G.~Zwicknagl: Phys. Rev. B {\bfseries 23} (1981) 5788.

\bibitem{ABSb}
J.~Hara and K.~Nagai: Prog. Theor. Phys. {\bfseries 76} (1986) 1237.

\bibitem{Hu}
C.~R. Hu: Phys. Rev. Lett. {\bfseries 72} (1994) 1526.

\bibitem{ABSR1}
S.~Kashiwaya and Y.~Tanaka: Rep. Prog. Phys. {\bfseries 63} (2000) 1641.

\bibitem{ABSR2}
T.~L{\"o}fwander, V.~S. Shumeiko, and G.~Wendin: Supercond. Sci. Technol.
  {\bfseries 14} (2001) R53.

\bibitem{YTK97}
M.~Yamashiro, Y.~Tanaka, and S.~Kashiwaya: Phys. Rev. B {\bfseries 56} (1997)
  7847.

\bibitem{YTK98}
M.~Yamashiro, Y.~Tanaka, Y.~Tanuma, and S.~Kashiwaya: J. Phys. Soc. Jpn.
  {\bfseries 67} (1998) 3224.

\bibitem{Yakovenko}
H.~Kwon, K.~Sengupta, and V.~Yakovenko: Eur. Phys. J. B {\bfseries 37} (2004)
  349.

\bibitem{Mao}
Z.~Mao, K.~Nelson, R.~Jin, Y.~Liu, and Y.~Maeno: Phys. Rev. Lett. {\bfseries
  87} (2001) 037003.

\bibitem{Kashiwaya11}
S.~Kashiwaya, H.~Kashiwaya, H.~Kambara, T.~Furuta, H.~Yaguchi, Y.~Tanaka, and
  Y.~Maeno: Phys. Rev. Lett. {\bfseries 107} (2011) 077003.

\bibitem{Maeno}
Y.~Maeno, H.~Hashimoto, K.~Yoshida, S.~Nishizaki, T.~Fujita, J.~G. Bednorz, and
  F.~Lichtenberg: Nature {\bfseries 372} (1994) 532.

\bibitem{Maeno2}
A.~P. Mackenzie and Y.~Maeno: Rev. Mod. Phys. {\bfseries 75} (2003) 657.

\bibitem{STYY11}
M.~Sato, Y.~Tanaka, K.~Yada, and T.~Yokoyama: Phys.\ Rev.\ B {\bfseries 83}
  (2011) 224511.

\bibitem{Volovik97}
G.E.Volovik: JETP Lett. {\bfseries 66} (1997) 522.

\bibitem{FMS01}
A.~Furusaki, M.~Matsumoto, and M.~Sigrist: Phys. Rev. B {\bfseries 64} (2001)
  054514.

\bibitem{tanaka12}
Y.~Tanaka, M.~Sato, and N.~Nagaosa: J. Phys. Soc. Jpn. {\bfseries 81} (2012)
  011013.

\bibitem{TYN09}
Y.~Tanaka, T.~Yokoyama, and N.~Nagaosa: Phys. Rev. Lett. {\bfseries 103} (2009)
  107002.

\bibitem{takami2014}
S.~Takami, K.~Yada, A.~Yamakage, M.~Sato, and Y.~Tanaka: J. Phys. Soc. Jpn.
  {\bfseries 83} (2014) 064705.

\bibitem{Reeg}
C.~Reeg and D.~L. Maslov: Phys. Rev. B {\bfseries 95} (2017) 205439.

\bibitem{Schnyder08}
A.~P. Schnyder, S.~Ryu, A.~Furusaki, and A.~W.~W. Ludwig: Phys. Rev. B
  {\bfseries 78} (2008) 195125.

\bibitem{James}
J.~J. He, J.~Wu, T.-P. Choy, X.-J. Liu, Y.~Tanaka, and K.~T. Law: Nature
  Communications {\bfseries 5} (2014) 3232.

\bibitem{QHZ10}
X.~L. Qi, T.~L. Hughes, and S.~C. Zhang: Phys. Rev. B {\bfseries 82} (2010)
  184516.

\bibitem{Ji}
W.~Ji and X.-G. Wen: Phys. Rev. Lett. {\bfseries 120} (2018) 107002.

\bibitem{Huang}
Y.~Huang, F.~Setiawan, and J.~D. Sau: Phys. Rev. B {\bfseries 97} (2018)
  100501.

\bibitem{He2017}
Q.~L. He, L.~Pan, A.~L. Stern, E.~C. Burks, X.~Che, G.~Yin, J.~Wang, B.~Lian,
  Q.~Zhou, E.~S. Choi, K.~Murata, X.~Kou, Z.~Chen, T.~Nie, Q.~Shao, Y.~Fan,
  S.-C. Zhang, K.~Liu, J.~Xia, and K.~L. Wang: Science {\bfseries 357} (2017)
  294.

\bibitem{Ii2011}
A.~Ii, K.~Yada, M.~Sato, and Y.~Tanaka: Phys. Rev. B {\bfseries 83} (2011)
  224524.

\bibitem{Ii2012}
A.~Ii, A.~Yamakage, K.~Yada, M.~Sato, and Y.~Tanaka: Phys. Rev. B {\bfseries
  86} (2012) 174512.

\bibitem{YamakagePhysicaE}
A.~Yamakage and M.~Sato: Physica E: Low-dimensional Systems and Nanostructures
  {\bfseries 55} (2014) 13 .

\bibitem{YTN09}
T.~Yokoyama, Y.~Tanaka, and N.~Nagaosa: Phys. Rev. Lett. {\bfseries 102} (2009)
  166801.

\end{thebibliography}

\end{document}